# Near-duplicate video detection featuring coupled temporal and perceptual visual structures and logical inference based matching

Mohammed Belkhatir [a,*], Bashar Tahayna [b,1]

[a] Faculty of Computer Science, University of Lyon, Campus de la Doua, 69622 Villeurbanne Cedex, France
[b] Faculty of Information Technology, Monash University, Sunway Campus, 46150, Malaysia

## ABSTRACT

We propose in this paper an architecture for near-duplicate video detection based on: (i) index and query signature based structures integrating temporal and perceptual visual features and (ii) a matching framework computing the logical inference between index and query documents. As far as indexing is concerned, instead of concatenating low-level visual features in high-dimensional spaces which results in curse of dimensionality and redundancy issues, we adopt a perceptual symbolic representation based on color and texture concepts. For matching, we propose to instantiate a retrieval model based on logical inference through the coupling of an *N*-gram sliding window process and theoretically-sound lattice-based structures. The techniques we cover are robust and insensitive to general video editing and/or degradation, making it ideal for re-broadcasted video search. Experiments are carried out on large quantities of video data collected from the TRECVID 02, 03 and 04 collections and real-world video broadcasts recorded from two German TV stations. An empirical comparison over two state-of-the-art dynamic programming techniques is encouraging and demonstrates the advantage and feasibility of our method.

## 1. Introduction

Near-duplicate video (NDV) detection in large multimedia collections is very important for digital rights management as well as for video retrieval applications. One crucial step for such task is to define a matching/mismatching measure between two video sequences. Extended research has been carried out to identify NDVs in video collections (Bertini, Bimbo, & Nunziati, 2006; Hoad & Zobel, 2003, 2006; Joly, Frèlicot, & Buisson, 2003; Joly, Buisson, & Frélicot, 2007; Vidal, Marzal, & Aibar, 1995; Zhou & Zhang, 2005). However, existing methods have substantial limitations: they are sensitive to the degradations of the video, expensive to compute, and mostly limited to the comparison of whole video clips. Moreover, much of the video content is distributed in a continuous stream that cannot be easily segmented for comparison, making these methods unsuitable for applications which are used by regulation authorities for continuous broadcast-streaming monitoring. During video editing, some inappropriate shots could be deleted and commercial breaks could be inserted. However, from the perspective of human perception, the initial and edited videos are still regarded as similar (Zhou & Zhang, 2005). Thus, in order to identify duplicates of a specific video, an efficient video matching and scoring framework is required for detecting similar or quasi-similar contents. Many of the existing matching models are not suitable for such a task since they either ignore the temporal dimension or simplify the query model. NDV detection requires models for video sequence-to-sequence matching incorporating the temporal order inherent in video data. For sequence matching to be meaningful, corresponding video contents shall be identified in a fixed chronological temporal order while ignoring all the in-between mismatching shots which are often artificially introduced in edited videos. To achieve this, many solutions are based on proposing to view a sequence of video frames as a string and use direct comparison between the sequence of features of the query and index videos. However, this method is computationally expensive and sensitive to changes that can occur during video editing. In order to reduce the computational cost, an alternative approach consists in computing a shot-based index structure viewed as a string and then apply string matching algorithms to solve the problem of shot alignment. The main programming paradigm used in the literature for computing sequence alignment, *dynamic programming*, has however some limitations. Its computational load is indeed affected by the number of shots and their duration. Furthermore, dynamic programming measures, in terms of edit distance, how mismatching two videos are rather than how similar they are.

In this paper, we propose a near-duplicate video detection framework based on signature-based index structures featuring perceptual visual attributes and a matching and scoring framework relying on logical inference. As far as indexing is concerned, the concatenation of low-level visual features (color, texture, etc.) in high-dimensional spaces traditionally results in curse of dimensionality and redundancy issues. Moreover, this usually requires normalization which may cause an undesirable distortion in the feature space. Indeed, since low-level visual features (color and texture) are of high dimensionality (typically in the order of $10^2$–$10^3$) and data in high-dimension spaces are sparse, it is necessary to gather enough observations to make sure that the estimation is viable. Consequently, it is crucial to consider the dimensionality reduction of the visual feature representation spaces. Moreover, contrary to the state-of-the-art approaches for dimensionality reduction (such as principal component analysis, multidimensional scaling, singular value decomposition) which are **opaque** (i.e. they operate dimensionality reduction of input spaces without making it possible to understand the signification of elements in the reduced feature space), our framework will itself be based on a **transparent** readable characterization. We propose to reduce the dimensionality of signal features by taking into account a perceptual symbolic representation of the visual features based on color and texture concepts. A matching framework relying on the logical inference between index and query documents is instantiated through an *N*-gram sliding window technique coupled with fast lattice-based processing. Near-duplicate videos are here defined as a set of matched pair-wise sequences, but with certain constraints that can be induced by frame rate conversions and editing, which abundantly exist in real-world applications.

Experimentally, we implement our theoretical proposition, detail the automatic characterization of the visual (color and texture) concepts and evaluate the prototype on 286 videos from the TRECVID 02, 03 and 04 corpora against two dynamic programming frameworks.

The remainder of this paper is organized as follows: Section 2 introduces the related work on NDV detection. Section 3 gives an overview of the proposed system architecture. Temporal video segmentation is detailed in Section 4. Signature-based indexing with duration, color and texture feature extraction is detailed in Section 5. Then in Section 6 we discuss the *N*-gram matching and scoring framework. Experimental results are reported in Section 7.

## 2. Related works

Many existing approaches support near-duplicate key-frame (image) detection such as presented in Ke, Sukthankar, and Huston (2004), Lowe (2003), Wu, Hauptmann, and Ngo (2007), Zhang and Chang (2004) and Zhao, Ngo, Tan, and Wu (2007). They cannot however be effectively applied to NDV detection due to the inherent characteristics of video documents and the limitation of the concerned techniques solely operating on the image dimension. Finding NDVs using shot-feature sequences requires a search method that is capable of efficiently comparing signatures to accurately locate similar sequences. In Mohan (1998), a representation of the video is produced by computing an ordinal signature for a reduced-intensity version of each frame; these are then concatenated to form a vector, which is used to determine similarity. For this, a sliding window is used to align the query and index feature vectors and the computation of an average distance between frames is proposed. However, direct comparison of ordinal signatures is computationally expensive, making this method unsuitable for large collections. Hoad and Zobel propose several methods to match video sequences. The first method identifies cuts or edits in the video and uses shot lengths as a video sequence descriptor (Hoad & Zobel, 2003). Dynamic programming is used to calculate the similarity between two strings in terms of the minimum number of insertion and deletion operations necessary to transform one video descriptor string into the other. This matching algorithm searches ten hours of the archive in one second, but the results are not always reliable, especially when the query has very few cuts. Zhou and Zhang (2005) propose two edit distance computations for near-duplicate detection. The first one is directly affected by the duration and the number of shots of the video. The second one is tested on videos either overlapping completely or non-overlapping, while ignoring queries that include partial overlapping or edited versions of the video. In Hoad and Zobel (2006), four techniques for generating index video representations are proposed: the shot-duration, color-shift, centroid-based, and combined methods. Authors use a dynamic programming-based matching algorithm to compute detection scores. They demonstrate video degradation by changes in video bit-rate, frame rate, and resolution. Their method based on combined features provides promising results. However, their experiment using shot-length features proves that the effectiveness of retrieval using dynamic programming is not on par with that of more computationally intensive approaches (e.g. those based on frame-by-frame comparison of visual features). We use an equivalent dynamic programming-based technique as a baseline for our experiments in Section 7.

For video scoring, dynamic programming-based methods are directly influenced by the shot duration features and the video length. Therefore, the performance and the true measurement of the edit distance can be degraded especially if the video has been edited and the shot durations modified by the insertion of commercials. Vidal et al. (1995) define the *Normalized Edit Distance* (NED) which solves this problem and could therefore be used. Unfortunately, the NED has a cubic complexity and is thus not suited for fast retrieval. This approach is used for comparison purposes in the experimental section.

Cheung & Zakhor address the problem of copy detection of video clips from the web (Cheung & Zakhor, 2000a,2000b). Individual frames are compared across video clips to match sequences, however ignoring temporal relationships. Hampapur et al. compare a variety of distance measures for video similarity measurement: intensity difference, color histogram intersection, gradient direction histogram intersection, partial Hausdorff distance between Canny edges, local edge representation (centroids of edge points within frame blocks), and the second order central moments of Canny edges within frames (Hampapur, Hyun, & Bolle, 2001). They then perform content-based copy detection using motion ordinal intensity signatures (Hampapur & Bolle, 2002). The latter are calculated by dividing the frame into square blocks, computing a motion vector for each of them, and counting the number of blocks in each quantized motion direction.

## 3. System architecture

Fig. 1 demonstrates the system architecture where a re-broadcasted video can be used as a query into the system. The query video is segregated by the segmentation unit into shots. Each shot is labeled wit its size in terms of number of frames. The signature extraction unit computes the signature of the input query consisting of visual and temporal features. The latter is then compared to the index video signatures by the video recognition unit that performs matching and scoring. As we are concerned with video clips that are derived from the same original source (with potential modifications, i.e. editing and/or adjunction of commercials), the system retrieves candidate videos based on a threshold featuring the mismatch percentage between query and index videos.

We detail the components of our video near-duplicate detection architecture as highlighted in Fig. 1:

1. A broadcasted video is captured and stored in a video corpus.
2. A corresponding index structure consisting of temporal, color and texture signatures is computed and itself stored in an index base.
3. A query video broadcast is processed by the video segmentation unit for abrupt and gradual edits detection.
4. The query video edit patterns are then labeled and its shots numbered.
5. The feature extraction unit highlights a query structure consisting of temporal, color and texture signatures.
6. The matching module compares video query and index signatures.

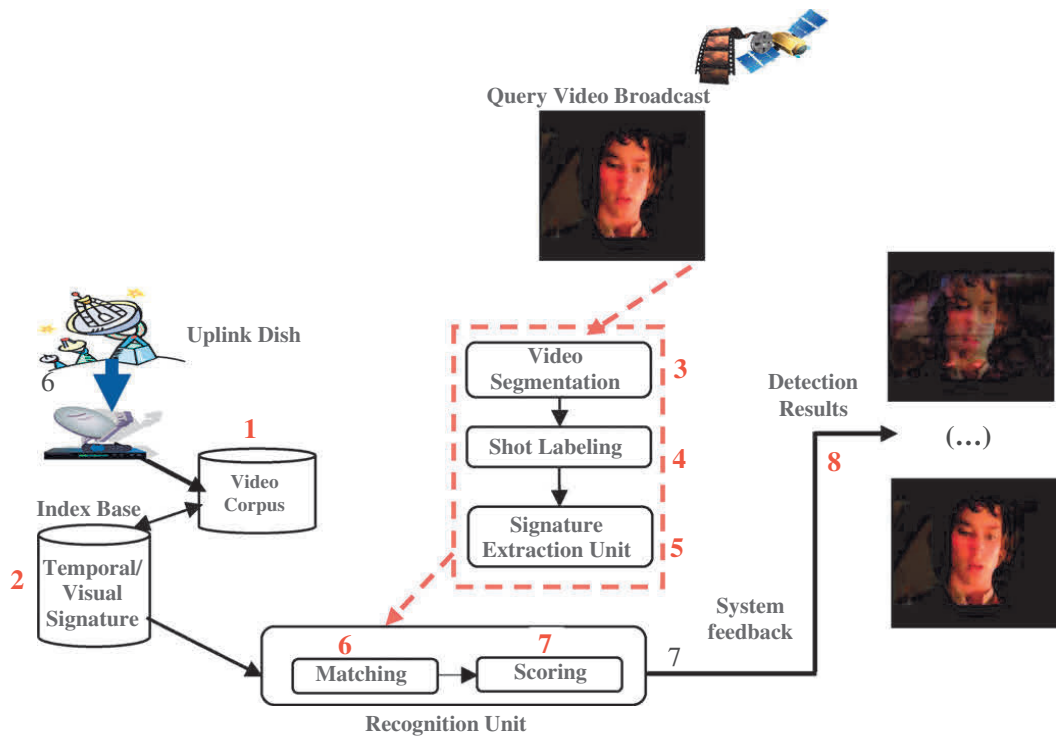

**Fig. 1.** System architecture for near-duplicate video detection.

7. The scoring unit is based on a ranking scheme to order index videos based on their relevance with respect to the query video.
8. The retrieval results are finally displayed through the system interface.

## 4. Temporal video segmentation

Segmentation aims at representing a video in its hierarchical structure whereby searching and accessing its information can be more effective. The first step in specifying a recognition system consists in segregating a video into its subunits. A segmentation algorithm shall partition a video robustly across encoding rates. An important source of information in visual content for segmentation is that of color. The simplest method is to use the frame difference to determine shot transitions. This method keeps track of the number of pixels transformed from a frame to the next frame. Color histograms are the most widely used color feature representations.

We make use of algorithms and techniques on both compressed and uncompressed video data. For the uncompressed domain (i.e., in our experimental corpus, 380 TV-show episodes recorded from two German TV stations) we apply various techniques to identify cuts, dissolves and transitions based on the difference of grey level histograms and energy of difference image as in Zhang, Kankanhalli, and Smoliar (1993), Arman, Hsu, and Chiu (1994), Zabih, Miller, and Mai (1995). On the compressed domain (i.e. the TRECVID data in MPEG-1 format), we consider the state-of-the-art method mentioned in Eickeler and Hesseler (2006) to detect the video abrupt changes only. The result of a video segmentation process is a signature which contains the frame temporal positions (in milliseconds with respect to the start of the video) to discriminate shots from each other. While computing the duration feature turns to be simple, the conversion between frame rates results in sensible variations with respect to the original signal, and in slight variations of the playback speed and the number of frames which constitute the video shot. To resolve the possible frame-rate variations between a video and an NDV, we use shot durations instead of number of frames. The algorithm computes the duration between the successive key-frames, and produces a signature which contains the temporal durations of shots. Notice that for more accuracy, we could use a normalized shot length, with respect to a given video length; however, we assume this is unknown for the real-time monitoring video stream application (here a differential offset could be used but this would cause several errors if an edited video copy with many new inserted commercials is considered).

## 5. Index and query representations

The choice of the video representative features and their integration within index and query structures is crucial for developing a successful matching and scoring framework since its performance and robustness largely depend in turn on the specific properties of the selected features and their integration. The matching between query and index video representations is then determined by some well-defined similarity measure in the feature space. In our architecture, the index and query video representations are computed at the shot level. It consists in a structure $s$ unifying three signature-based representations, $s = (s_{duration}, s_{color}, s_{texture})$ where $s_{duration}$ is characterized by a vector structure featuring the video temporal characterizations presented in Section 5.1, $s_{color}$ and $s_{texture}$ are represented by vector structures respectively characterizing the visual color and texture descriptions and detailed in Sections 5.2 and 5.3. The visual information is here described by perceptual symbolic descriptors, i.e. the color and texture concepts, through processes mapping them to the extracted low-level features.

### 5.1. Duration-based features

The duration of a video shot is here used as an index descriptor. Although being a crude description since it does not carry any information about the content of the shot, it is very compact, trivial to compute and its generation is computationally less expensive than the characterization of visual features. While computing the duration feature turns to be simple, the conversion between frame rates results in sensible variations with respect to the original signal, and in slight variations of the playback speed and the number of frames which constitute the video shot. To resolve the possible frame-rate variations between a video and an NDV, we use shot durations instead of number of frames. The algorithm computes the duration between the successive key-frames of a shot, and stores them in the signature $s_{duration}$ with a reference to the video they belong to. The following is an example of a typical shot-duration signature (in milliseconds), which represents a two-minute video clip using only nine scalar values: [1170, 2610, 1020, 8320, 19640, 20220, 23230, 27310, 16480].

### 5.2. Color-based features

Our representation of color information is guided by the research carried out in color naming and categorization. Inspired by Berlin and Kay, works have revolved around stressing a step of correspondence between color stimuli and 'basic color terms' (Berlin & Kay, 1991) which they characterize by the following properties: their application is not restricted to a given object class, i.e. the color characterized by the term "olive color" is not valid; they cannot be interpreted conjointly with object parts, i.e. "the maple leaf color" is not a valid color; their interpretation does not overlap with the interpretation of other color terms and finally they are psychologically meaningful. Further works proposed in Gong and Al (1996) consist of an experimental validation of the 'basic color term' notion in the HVC perceptual color space. The latter belongs to the category of user-oriented color spaces (as opposed to material-oriented spaces such as RGB), i.e. spaces which define color as being perceived by a human through tonality (describing the color wavelength), saturation (characterizing the quantity of white light in the color spectral composition) and brightness (related to color intensity). Given a series of perceptual evaluations and observations, eleven color concepts: black ($cc_1$), blue ($cc_2$), cyan ($cc_3$), green ($cc_4$), grey ($cc_5$), orange ($cc_6$), purple ($cc_7$), red ($cc_8$), skin ($cc_9$), white ($cc_{10}$), yellow ($cc_{11}$) are highlighted, each described by tonality, brightness and saturation values.

Characterizing the aforementioned perceptual color concepts involves algorithmically transforming the extracted low-level features specified in the RGB space (primary step for low-level color extraction) to tonality, brightness and saturation values in the perceptually uniform HVC space. This step is detailed in the experimental Section 8.1.1.

A signature $s_{colour}$ with eleven elements corresponding to the aforementioned color concepts features the color distribution of a video shot. The value $s_{color}[i]$ ($i \in [1, 11]$) is the pixel percentage of color concept $cc_i$ within the corresponding shot. For example, the vector ⟨black:20, blue:0, cyan:0, green:17, grey:0, orange:12, purple:0, red:0, skin:0, white:30, yellow:21⟩ corresponds to the color distribution of a shot with 20% of black, 17% of green, 12% of orange, 30% of white and 21% of yellow.

*5.3. Texture-based features*

Although several works have proposed the identification of low-level features and the development of algorithms and techniques for texture computation, few attempts have been made to propose an ontology for texture symbolic characterization and naming. In Bhushan and Al (1997), a texture lexicon consisting of eleven high-level texture concepts is proposed as a basis for perceptual texture classification. In each of these concepts, a texture concept which best describes the nature of the characterized texture is proposed.

The first texture category $C_1$ gathers textures with random three dimensional imperfections and is characterized by the texture concept bumpy ($tc_1$). $C_9$ comprises textures exhibiting random linear orientation and is represented by the texture concept cracked ($tc_2$). $C_3$ gathers textures that do not present any structure nor any dominant orientation. It is represented by the texture concept disordered ($tc_3$). $C_4$ gathers structured textures with a weave-like structure. Textures in $C_4$ differ from textures in category $C_3$ as they present a certain amount of variation and randomness. This category is represented by the texture concept interlaced ($tc_4$). $C_5$ consists of linearly oriented textures (the orientation is along a straight line). It is represented by the texture concept lined ($tc_5$). The category $C_6$ consists of textures characterized by the texture concept marbled ($tc_6$). The seventh texture category $C_7$ consists of texture with two directional characteristic features, combined to form a weave. It is represented by the texture concept netlike ($tc_7$). $C_8$ regroups textures presenting some disfigurement. It is characterized by the texture concept smeared ($tc_8$) denoting negative aesthetics. Texture category $C_9$ consists of textures with representative features being small, blob-like and scattered over a plane. It is characterized by the texture concept spotted ($tc_9$). The tenth texture category $C_{10}$ refers to uniform ($tc_{10}$) textures. The nature of the repetition is not specified in this category, which was the case for textures in category $C_7$. Texture category $C_{11}$ consists of circularly oriented textures. It is represented by the texture concept whirly ($tc_{11}$). Low-level texture features are mapped to texture concepts and the latter are assigned posterior recognition probabilities through a support vector machine based framework detailed in Section 8.1.2.

A signature $s_{texture}$ with eleven elements corresponding to the aforementioned texture concepts features the texture distribution of a video shot. The value $s_{texture}[i]$ ($i \in [1, 11]$) is a boolean translating that texture concept $tc_i$ characterizes the corresponding shot. For example, the vector ⟨bumpy:1, cracked:1, disordered:0, interlaced:0, lined:0, marbled:0, netlike:0, smeared:1, spotted:0, uniform:0, whirly:0⟩ corresponds to a distribution with bumpy, cracked and smeared textures.

## 6. A matching and scoring framework based on logical inference

The matching framework is based on a logical inference model where the relevance of an index video $I$ with respect to a query $Q$ is given by the product of the exhaustivity measure $E$ and the specificity measure $S$:

$$\text{Relevance}(I, Q) = E(I \to Q) \times S(Q \to I)$$

We detail the exhaustivity and specificity functions below.

*6.1. Exhaustivity*

Exhaustivity measures to which extent the video document satisfies the query through a logical implication and its instantiation consists in two operations. It first checks that all elements described within the query representation are also elements of the index representation. For this, we use a method combining an $N$-gram sliding window framework operating on the temporal signature with lattice-based projection on the visual color and texture signatures described as follows:

1. Consider $N$ successive sliding windows, each corresponding to a given shot, applied on the query and index videos (the value $N$ is determined experimentally).

2. Apply pair-wise temporal signature comparison of the query and index adjacent shots in the sliding windows. For equal temporal signatures, perform lattice projection for the color and texture signatures to evaluate their correspondence (the organization of the lattice is described in Section 6.2). The final steps consist of counting the number of matches, reserving the current offset and the corresponding number of matches.
3. Slide the index windows $S$ steps forward, repeat step 2.
4. For each complete scan of the index video, retrieve the offset such that the number of matched shots is maximal ($max(number\_of\_matched\_shots)$) and the corresponding position in the query video.
5. Slide the query windows $S$ steps forward.
6. Repeat from step 2 until the query video is completely scanned. Assign the best starting position and offset for the maximum number of matches and their offset.
7. Starting from this position, count the number of equal pairs and assign a score to the index video.

The latter consists of an estimation of its relevance with respect to the query, which corresponds to the quantitative evaluation of their similarity. It is given by the exhaustivity value between query signature $Q$ and index signature $I$:

$$SV(Q,I) = \sum_{s_{texture\_q} \text{ of } Q, s_{texture\_i} \text{ of } I} Path\_Tex(s_{texture\_q}, s_{texture\_i}) + \sum_{s_{color\_q} \text{ of } Q, s_{color\_i} \text{ of } I} Path\_Col(s_{color\_q}, s_{color\_i})$$

The $Cpt\_Match$ function is the Kullback–Leibler divergence between the probabilities of:

(i) texture concepts in the query texture signature $s_{texture\_q}$ (which are themselves certain, i.e. their probability is equal to 1) and texture concepts in the matching index texture signature $s_{texture\_i}$ (their probability is assigned by the learning-based framework presented in Section 8.1.2).
(ii) color concepts in the query color signature $s_{color\_q}$ (which are themselves certain, i.e. their probability is equal to 1) and color concepts in the matching index color signature $s_{color\_i}$ (their probability is defined by the ratio of pixels corresponding to a given concept over the total number of pixels within the corresponding frame).

Fig. 2 illustrates the $N$-gram sliding window technique for query and index video matching and scoring.

Moreover, lattices organizing color and texture signatures are defined by mathematical partial orders and consequently not stored in memory, which avoids traversing complex structures at retrieval. We detail in Section 7 the organization of signature lattices for effective query processing.

*6.2. Specificity*

Specificity itself measures the importance of the query themes within an index video, it is given by the value of $S(Q \rightarrow I)$. The specificity function takes into account the importance of the query elements within the index document. We assume that a user expects that detected videos are strictly restricted to visual elements related to those in the query video. If not, we say that the video document 'degrades' the query. Lattices of texture and color signatures (cf. Section 7) take into account the query degradation phenomenon by relating more closely texture signatures with the most common number of texture concepts and color signatures characterizing similar color distributions. Therefore, the evaluation of the query degradation is formally mapped to a path length evaluation problem in these lattices. The $Path\_Tex$ and $Path\_Col$ functions compute path lengths in lattices between matching index and query signatures.

The specificity value measures the importance of the query themes within the index document by minimizing path lengths between texture and color signatures of a query signature $Q$ and those of an index signature $I$:

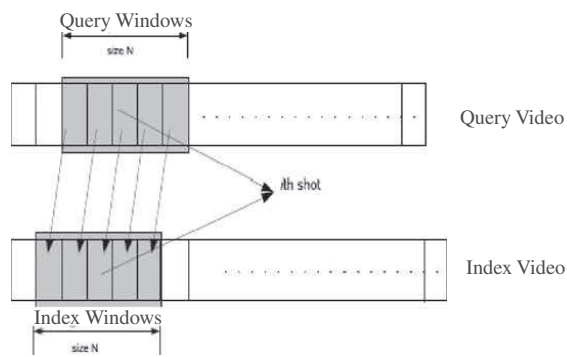

**Fig. 2.** $N$-gram sliding window technique for query and index video matching and scoring.

$$SV(Q,I) = \sum_{S_{\text{texture\_q}} \text{ of } Q, S_{\text{texture\_i}} \text{ of } I} Path\_Tex(S_{\text{texture\_q}}, S_{\text{texture\_i}}) + \sum_{S_{\text{color\_q}} \text{ of } Q, S_{\text{color\_i}} \text{ of } I} Path\_Col(S_{\text{color\_q}}, S_{\text{color\_i}})$$

To evaluate path lengths in signature lattices, it is important to note that no computationally-heavy lattice traversal is performed. Indeed, path lengths are computed on the fly through explicit mathematical relations. We specify next the organization of lattices processing texture and color signatures.

## 7. Organization of lattices for fast matching

### 7.1. Processing texture signatures

#### 7.1.1. Lattice description

The texture distribution of a video shot featuring perceptual textures 'bumpy' and 'cracked' is translated by the texture signature ⟨bumpy:1, cracked:1, disordered:0, interlaced:0, lined:0, marbled:0, netlike:0, smeared:0, spotted:0, uniform:0, whirly:0⟩. This query *signature* is then related to its equivalent index signature as highlighted in Fig. 3. The most relevant index shots have bumpy and cracked roads, i.e. they are characterized only by perceptual textures featured in the query signature. This texture distribution is represented by the highlighted texture index signature ($t_1$) in Fig. 3. Other shots have a texture distribution with at least one additional texture concept not featured in the query (called secondary). In the lattice, texture index signatures representing such distributions are descendants of $t_1$. Formally, we define a partial order in this lattice noted $\leqslant_{\text{text}}$ by:

$\forall a, b$ index texture signatures, $a \leqslant_{\text{text}} b \iff [a = \bot_{\text{text}} \vee b = \top_{\text{text}}] \vee [\neg \exists k \in [1, 11] / b_{[k]} = 1 \wedge a_{[k]} = 0]$.

#### 7.1.2. Path Lengths

In this lattice, we furthermore derive the path length between the minimum element $\bot_{\text{text}}$ and an index texture signature $t$ characterized by the number $n_t$ of its null components. Let us note that the number of null components of the minimum element is equal to 0.

We prove that this path length is equal to $\mathbf{11 - n_t}$ by induction on the number $n_t$ of the null components of $t$:

- For $n_t = 0$, $t$ is the minimum element $\bot_{\text{text}}$. The property is verified, the path length between an element of the lattice and itself being 0.
- We assume that the property is verified at rank $i$.
- At rank $i + 1$, we consider an index texture signature $t_1$ with a number of zero components equal to $n_{t_1} = i + 1$ (the number of non-zero elements is therefore equal to $11 - (i + 1)$). This element is the parent of an index texture signature $t_2$ with a number of null components equal to $i$. The path length between these two elements is equal to 1. The path length between $t_2$ and the minimum element $\bot_{\text{text}}$ is by recurrence hypothesis equal to $11 - (11 - i)$, i.e. $i$. Consequently, the path length between $t_1$ and $\bot_{\text{text}}$ is equal to $i + 1 = 11 - (11 - (i + 1))$. The property is therefore verified at rank $i + 1$.

### 7.2. Processing color signatures

There are two lattices processing color signatures taking into account the quantification conveyed by these structures: the *At Most* and *At Least* lattices.

The color distribution of a video query-by-example shot is first translated into a color signature. For example, a color distribution featuring 50% of black is translated into the color signature ⟨black:50, blue:0, cyan:0, green:0, grey:0, orange:0, purple:0, red:0, skin:0, white:0, yellow:0⟩.

#### 7.2.1. 'At Most' lattice

This signature is processed by the At Most lattice such that the most relevant index shots consist of key-frames with 50% of black and a remaining proportion uniformly distributed between the color concepts not represented in the query (5% each in our example), called secondary. Other shots consist of key-frames with a color distribution that includes less than 50% of black, the 'remaining' proportion $p$ being in the best cases uniformly distributed between the secondary color concepts.

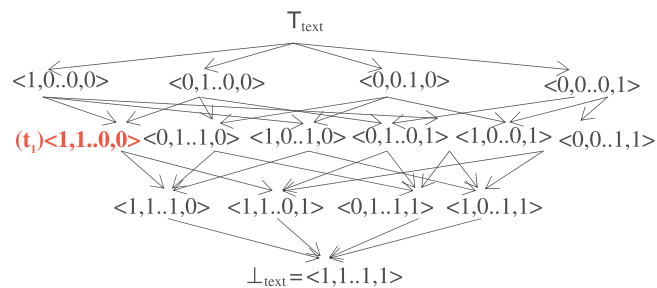

Fig. 3. Lattice processing *index* texture signatures.

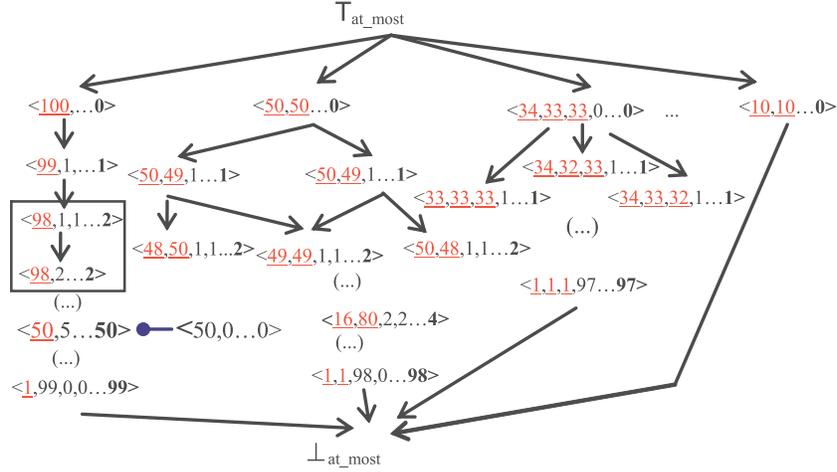

Fig. 4. *At Most* lattice processing color signatures.

The organization of the lattice takes into account **dominant** color concepts (i.e. concepts mentioned in a query as they have a higher importance in the ordering process of signatures within the lattice). For this the color signature is modified to include one additional dimension corresponding to a component summing pixel percentages of secondary color concepts noted $\sum_{at\_most}$. This novel structure, supported by a vector $s_{at\_most}$, is called color signature with dominant $d_{at\_most}$ where $d_{at\_most}$ is the set of dominant color concepts. The $s_{at\_most}[i]_{i\in[1,12]}$ values such that $c_i \in d_{at\_most}$ are the maximum pixel percentages of dominant color concepts and the $s_{at\_most}[j]_{j\in[1,12]}$ such that $j \neq i$ correspond to the pixel percentages of secondary color concepts ranked in ascending order. By construction, $\sum_{at\_most}$ is the maximum value among the $s_{at\_most}[j]_{j\in[1,12]}$. The query color signature ⟨black:50, blue:0, cyan:0, green:0, grey:0, orange:0, purple:0, red:0, skin:0, white:0, yellow:0⟩ is related to its equivalent color signature with dominant {black}:⟨50, 5, 5, 5, 5, 5, 5, 5, 5, 5, 5, 50⟩ as highlighted in Fig. 4.

Formally, sub-lattices of signatures with dominant $d_{at\ most}$ are partially ordered by $\leqslant_{at\ most}$. This partial relation takes into account the set $d_{at\ most}$ of dominant color concepts:

$\forall\ a, b$ signatures with dominant $d_{at\ most}$, $a \leqslant_{at\ most} b \iff [a = \bot_{at\ most} \vee b = T_{at\ most}] \vee [\forall j \in [1,12]/C_j \in d_{at\ most}, 1 \leqslant a_{[j]} \leqslant b_{[j]}]$

Sub-lattices of signatures with components corresponding to dominant color concepts being equal (framed structure in Fig. 4) are partially ordered by $\leqslant_{at\ most\_eq}$:

$\forall\ a, b$ signatures with dominant $d_{at\ most}$ having components that correspond to dominant color concepts being equal, $a \leqslant_{at\ most\_eq} b \iff \forall j, k \in [1,12]/C_j \notin d_{at\ most} \wedge C_k \notin d_{at\ most}, \sum_{j,k}|b_{[j]} - b_{[k]}| \leqslant \sum_{j,k}|a_{[j]} - a_{[k]}|$

### 7.2.2. 'At Least' lattice

This signature is also processed by the At Least lattice such that the most relevant index shots consist of key-frames with 50% of black and a remaining proportion uniformly distributed between the color concepts not characterized in the query (5% each in our example), called secondary. Other shots consist of key-frames with a color distribution that includes more than 50% of black, the 'remaining' proportion $p$ being in the best cases uniformly distributed between the secondary color concepts.

The organization of the lattice also takes into account **dominant** color concepts. For this the color signature is modified to include one additional dimension corresponding to a component summing pixel percentages of secondary color concepts noted $\sum_{at\_least}$. This novel structure, supported by a vector $s_{at\_least}$, is called color signature with dominant $d_{at\_least}$ where $d_{at\_least}$ is the set of dominant color concepts. The $s_{at\_least}[i]_{i\in[1,12]}$ values such that $c_i \in d_{at\_least}$ are the maximum pixel percentages of dominant color concepts and the $s_{at\_least}[j]_{j\in[1,12]}$ such that $j \neq i$ correspond to the pixel percentages of secondary color concepts ranked in ascending order. By construction, $\sum$ is the maximum value among the $s_{at\_least}[j]_{j\in[1,12]}$. The query color signature ⟨black:50, blue:0, cyan:0, green:0, grey:0, orange:0, purple:0, red:0, skin:0, white:0, yellow:0⟩ is related to its equivalent color signature with dominant {black}:⟨50, 5, 5, 5, 5, 5, 5, 5, 5, 5, 5, 50⟩ as highlighted in Fig. 5.

Formally, sub-lattices of signatures with dominant $d_{at\_least}$ are partially ordered by $\leqslant_{at\ most}$. This partial relation takes into account the set $d_{at\_least}$ of dominant color concepts:

$\forall\ a, b$ signatures with dominant $d_{at\_least}$, $a \leqslant_{at\_least} b \iff [a = \bot_{at\_least} \vee b = T_{at\_least}] \vee [\forall j \in [1,12]/C_j \in d_{at\_least}, b_{[j]} \leqslant a_{[j]} \leqslant 100]$.

Sub-lattices of signatures with components corresponding to dominant color concepts being equal (framed structure in Fig. 5) are partially ordered by $\leqslant_{at\_least\_eq}$:

$\forall\ a, b$ signatures with dominant $d_{at\_least}$ having components that correspond to dominant color concepts being equal, $a \leqslant_{at\_least\_eq} b \iff \forall j, k \in [1,12]/C_j \notin d_{at\_least} \wedge C_k \notin d_{at\_least}, \sum_{j,k}|b_{[j]} - b_{[k]}| \leqslant \sum_{j,k}|a_{[j]} - a_{[k]}|$

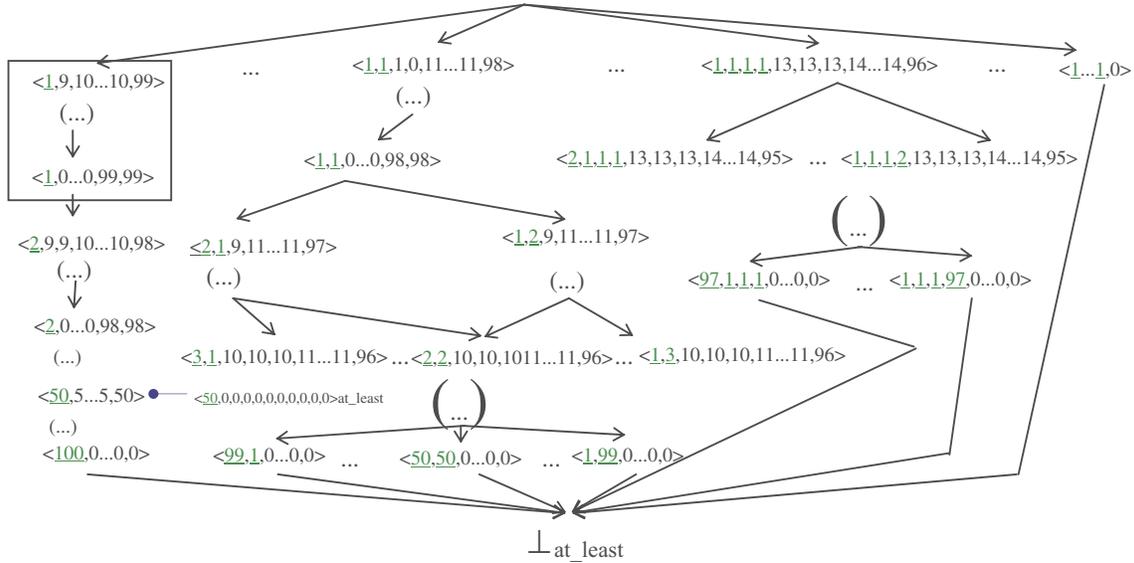

Fig. 5. *At Least* lattice processing color signatures.

### 7.2.3. Path lengths

To determine path lengths in the At Most lattice, we first characterize the depth of a sub-lattice structure (such as the framed structure in Fig. 4) which consists of elements corresponding to dominant color concepts being equal.

It is equal to the number of partitions Part($p, n_c$–Card($d_{at\_most}$)) of the integer $p$ characterizing the pixel proportion of the $11 - \text{Card}(d_{at\_most})$ secondary color concepts as a sum from one to $11 - \text{Card}(d_{at\_most})$ positive integers. To each of these partitions is associated a color signature with dominant $d_{at\_most}$. In the case of the framed structure of Fig. 4, the depth is equal to 2. Indeed, the proportion of the 10 secondary color concepts is equal to 2. The number of decompositions Part(2, 10) of the integer 2 as a sum from one to 10 positive integers equals 2. Indeed:

$$2 = 2 + 0 + \cdots + 0$$

$$2 = 1 + 1 + \cdots + 0$$

The position of the integers in the decomposition within the color signature does not carry any influence on determining the structure depth.

From this result, we can therefore establish the path length in the sub-lattice from the minimum element $\perp_{am}$ to an element with a proportion of dominant color concepts equal to $P$ and for which the pixel proportion of secondary color concepts is assumed uniformly distributed. It is equal to: $[\sum_{i=1,\ldots,P} \text{Part}(i, 11 - \text{Card}(d_{at\_most}))] + 1$.

With a similar reasoning in the *At Least* lattice, the path length in the sub-lattice from the minimum element $\perp_{am}$ to an element with a proportion of dominant color concepts equal to $P$ and for which the pixel proportion of secondary color concepts is assumed uniformly distributed. It is equal to: $[\sum_{i=1,\ldots,P} \text{Part}(i, 11 - \text{Card}(d_{at\_least}))] + 1$.

## 8. Experiments

We discuss in the experimental section the automatic characterization of the perceptual visual features and then propose the evaluation of the full framework for near-duplicate video detection.

### 8.1. Automatic characterization of the perceptual visual features

#### 8.1.1. Highlighting color concepts

After a first step of low-level color extraction in the RGB space for each pixel of a video frame, we set up a transformation process for characterizing this information in the HVC space.

Indeed, the use of the RGB color space firsthand is inefficient since the perceptual similarity between color pairs is not taken into account. Consequently, the color information is conveyed in the HVC perceptual space, which is moreover uniform.

Components H, V and C correspond respectively to the values of tonality, luminosity and saturation. They are then mapped to the eleven color concepts introduced in Section 5.2.

We iterate this process for all pixels and finally obtain the pixel percentage corresponding to each color concept for the frame processed. This information is then computed for all frames of the considered video shot and averaged. The resulting data constitute a vector structure consisting of eleven dimensions with each dimension representing the pixel percentage for a given color concept.

#### 8.1.2. Characterization of texture concepts

*8.1.2.1. Low-level computational extraction.* The study of texture in computer vision has lead to the development of several computational models for texture analysis used in several content-based image retrieval architectures. However, these texture extraction frameworks mostly fail to capture aspects related to human perception. Therefore, we propose a solution specifying a computational framework for texture extraction which is the closest approximation of the human visual system. The action of the visual cortex, where an object is decomposed into several primitives by the filtering of cortical neurons sensitive to several frequencies and orientations of the stimuli, is simulated by a bank of Gabor filters. We focus on computational texture extraction at the level of a video frame and characterize it by its Gabor energy distribution within seven spatial frequencies covering the whole spectral domain and seven angular orientations. It is then represented by a 49-dimension vector, with each dimension corresponding to a Gabor energy.

*8.1.2.2. Support vector machine based mapping.* The eleven high-level texture concepts presented in Section 5.3, foundation of our framework for texture perceptual characterization are automatically mapped to the 49-dimension vectors of Gabor energies through support vector machines. In order to determine which texture concept is associated with a given texture distribution, we have a set of points $\{x_1, \ldots, x_i \ldots\}$ in an $n$-dimensional input space $S^n$ of low-level texture features (here $n$ = 49), a set of labels $\{y_1, \ldots, y_i \ldots\}$ such that the $y_i$ value equals 1 if $x_i$ corresponds to texture concept $tc_i$ $c_{sem}[i]$ and $-1$ otherwise. The goal is to determine a function $f: S^n \rightarrow \{\pm 1\}$ which associates each low-level texture feature with its corresponding concept. For this, we consider support vector machines which, for separable problems, are based on algorithms highlighting the unique optimal hyperplane discriminating the data among the class of hyperplanes. The learning process consists in maximizing a function which considers the distance between each training data and class borders. The optimal position of a class border is obtained as a linear combination of training data within the border neighborhood: they are called support vectors. The latter play a crucial role in the learning process. In the case of non-linearly separable problems, projection kernels are used and support vector machines are then based on the resolution of the following optimization problem: $\min_{w,b,\phi} \frac{1}{2} w^T w + C \sum_{i=1}^{l} \phi_i$ subject to $y_i(w^T \psi(x_i) + b) \geq 1 - \phi_i$ and $\phi_i \geq 0$.

Here, training vector $x_i$ is set in correspondence in a space of higher dimension (sometimes infinite) through the function $\psi$. Support vector machines then determine a separating linear hyperplane in this space. $K(x_i, x_j) = \psi(x_i)^T \psi(x_j)$ is the projection kernel and $C$ the penalty parameter of the error term. Among the possible kernels (linear, polynomial, radial basis function, sigmoid, etc.), we choose the radial basis function: $K(x_i, x_j) = \exp(-\gamma \|x_i - x_j\|^2), \gamma > 0$ where $\gamma$ is a kernel parameter. It is traditionally used in the case of non-linearity between the class labels and the input attributes.

We adopt the "one-against-rest" classification strategy to optimize inter-class separation. A posterior recognition probability is assigned for the classification through the use of a logistic classifier maximizing the likelihood of the classified training data.

*8.1.2.3. Testing.* The texture dataset consists of 10,000 texture images, each of the 11 high-level texture concepts being represented by 900 to 1000 texture images. We propose in Fig. 6 example images for each of the specified texture concepts.

To determine the performance of the mapping, we first use $v$-fold cross-validation. The training set is divided into $v$ subsets of equal size. Sequentially, one subset is tested using the classifier trained on the remaining $v - 1$ subsets. Thus, each instance of the whole training set is predicted once so the cross-validation accuracy is the percentage of data which are correctly classified. This procedure prevents overfitting. Then, we apply a grid-search procedure to find the optimal parameters achieving the best cross-validation accuracy. For each of the 11 high-level texture concepts, the best cross-validation rate is given in Table 1.

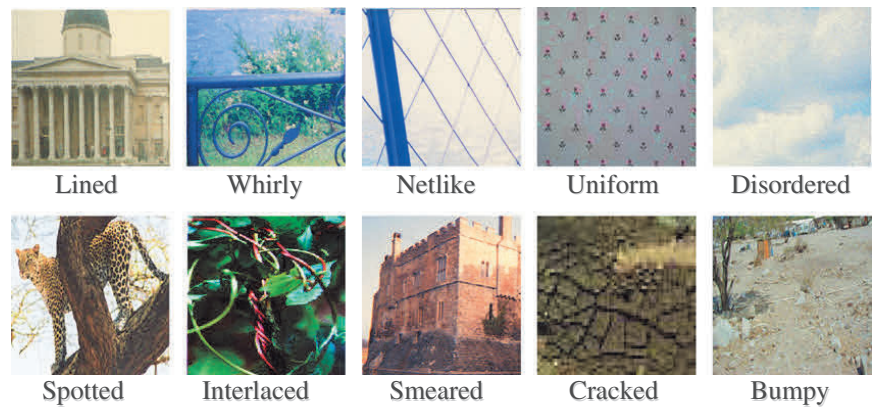

Fig. 6. Correspondence between texture images and texture concepts.

**Table 1**
Cross-validation percentages for texture concepts.

| Concept | Bumpy | Cracked | Disordered | Interlaced | Lined | Marble | Netlike | Smeared | Spotted | Uniform | Whirly |
|---|---|---|---|---|---|---|---|---|---|---|---|
| % | 83.7 | 85.2 | 88.9 | 91.9 | 94.5 | 98 | 86.8 | 83.4 | 90 | 97.3 | 81.4 |

**Table 2**
Parameter settings of the *N*-gram based detection framework.

| Query type | Window size (*N*) | Number of steps (*S*) |
|---|---|---|
| Re-broadcasted versions of C1 videos (Q1) | 200 | 20 |
| C1 videos edited with random introductions of commercials (Q2) | min(ShotsNumber(query, index)) | 1 |
| C2 videos edited with random introductions of commercials (Q3) | min(ShotsNumber(query, index)) | 1 |
| Short-length query videos (Q4) | min(ShotsNumber(query, index)) | 1 |

### 8.2. Evaluation of the near-duplicate video architecture

In this section we explore the effectiveness of the proposed *N*-gram sliding window technique for video matching and scoring compared to two state-of-the-art dynamic programming approaches with two experiments. The first experiment tests the effectiveness of the integrated temporal and visual index signatures and logical inference based matching and scoring framework on long query videos of different nature with multiple edited contents. The second experiment tests our architecture on short-length query videos, i.e. segments of videos in the index corpus.

#### 8.2.1. Index, query collections and parameter settings

Both experiments are conducted on two large corpora. The first, **C1**, is comprised of 176, 57 and 53 videos gathered from the TRECVID 02, 03, and 04 collections respectively (with a total of 286 videos). The video lengths in this data set are comprised between 3 and 37 minutes. The second experimental corpus, **C2**, consists of 180 h of video data of size 360 GB. It contains 380 episodes (and their near-duplicate versions) of three different television series: MacGyver (50 episodes), Good Times Bad Times (230 episodes) and Home Improvement (100 episodes), recorded from two German TV stations. The durations of videos in $C_2$ are almost uniform and range between 57 and 61 minutes at most. These have different artificial effects and different logos for the same broadcasted program by the two TV stations. The sizes of the frames are different for the two TV stations as well as their saturation and brightness. The rebroadcasted episodes (from both TV stations) have different breaking news banners and bottom banners as well as additional flying boxes different from the original ones. Three query collections are proposed for the first experiment:

- Query set Q1 consists of re-broadcasted versions of the videos in corpus C1 with different frame rates (29.97 fps vs. 25 fps for the initial version of the video in the index corpus).
- Query set Q2 consists of edited versions of the videos in corpus C1 with artificial random introduction of commercials.
- Query set Q3 consists of edited versions of the videos in corpus C2 with artificial random introduction of commercials.

In order to simulate the actual TV standards, these commercials are chosen with random size ranging in length from a few seconds to several minutes and covering at most 20% of the video length.

A fourth query set Q4, used in the second experiment, is comprised of short-length query videos as segments of videos in the index corpus. We furthermore propose in Table 2 the parameter settings used for the *N*-gram based detection framework.

#### 8.2.2. Experimental results

In both experiments, we compare our system to a state of the art system based on edit distance computation and representative of approaches in Bertini et al. (2006), Yeh and Cheng (2008). We also compare it to a system based on normalized edit distance implementation as described in Vidal et al. (1995), Arslan and Egecioglu (1999).

For the recognition evaluation, each video is labeled with a unique ID in order to compute the precision and recall of the system as well as the number of correct recognitions and falsely retrieved videos. The precision is the percentage of relevant videos in relation to the number of videos recognized. The recall is the percentage of the total relevant videos in the dataset that are recognized by the system.

In the first experiment, we compare the three systems on long query videos of different nature with multiple edited contents. For this, we use query sets Q1, Q2 and Q3 consisting each of 190, 250 and 250 video queries respectively. The results in Fig. 7 describe the recognition performance in terms of precision and recall metrics. The proposed technique based on logical inference matching outperforms the baseline frameworks in both precision and recall. We notice that even for high recall values (around 0.9) quantitatively measuring the propensity of the system to return all relevant index videos for a given query video, the precision is itself high and greater than 0.8. The latter quantitatively measures the propensity of the system to return only relevant index videos for a given query video. This is further corroborated by the experimental figures in Fig. 8 where the recognition performance is measured by the number of correct and falsely retrieved videos for the three compared systems. The mean average precision for all queries was computed for each compared system as: $\text{MAP}_{system} = \frac{\sum_{q=1}^{nQ_1} AveP(q)}{nQ_1} + \frac{\sum_{q=1}^{nQ_2} AveP(q)}{nQ_2} + \frac{\sum_{q=1}^{nQ_3} AveP(q)}{nQ_3}$ where $nQ_1$, $nQ_2$ and $nQ_3$ are the number of queries in each data set Q1, Q2 and Q3 respectively. The results obtained for the three compared systems are as follows: $\text{MAP}_{log\_infer}$ = 0.839, $\text{MAP}_{edit\_dist}$ = 0.754, $\text{MAP}_{norm\_edit\_dist}$ = 0.783. The average precision of our framework based on logical inference matching is approximately 11.3% higher over the average precision of the edit distance based framework and approximately 7.2% higher over the average precision of the system based on normalized edit distance.

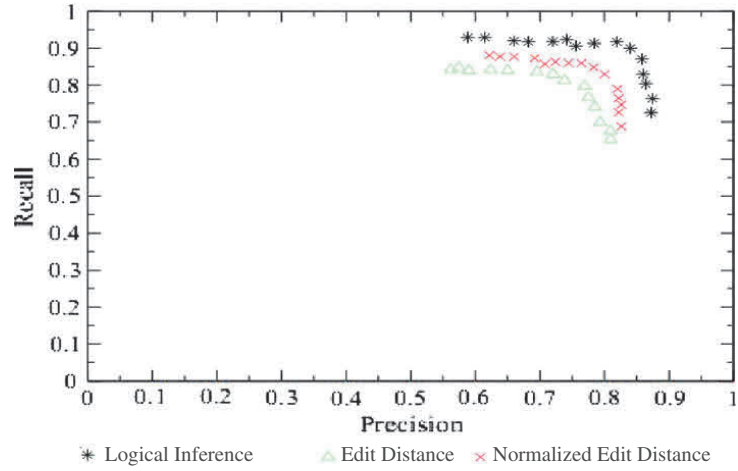

Fig. 7. Precision-recall with Q1, Q2 and Q3 query sets.

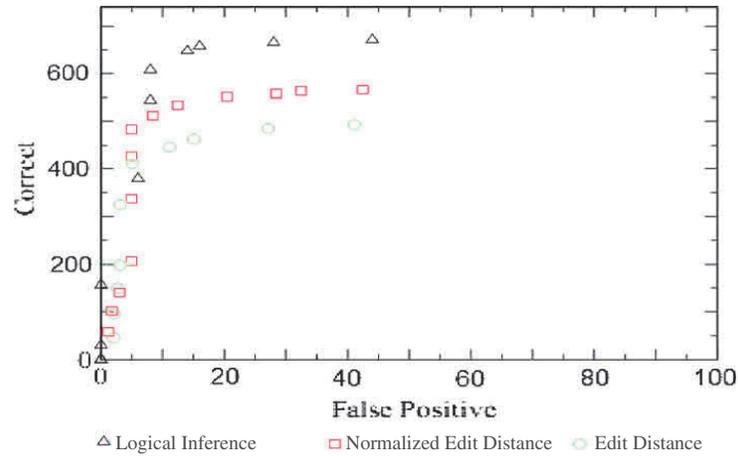

Fig. 8. Correct detection and false positive on query sets Q1, Q2 and Q3.

In the second experiment, we compare the effectiveness of the proposed framework compared to the dynamic programming methods on short-length queries. Query set Q4, used for this experiment, is further partitioned into three sets of 120 queries, each comprised of 5-second, 15-second and 2-minute query videos respectively. The results in Fig. 9 describe the recognition performance in terms of precision and recall. We notice that the overall performance declines for all three compared systems but as far as our system based on logical inference matching is concerned, even for high recall values (here around 0.8), the precision is itself high around 0.8.

We compute the MAP as in the previous experiment with the logical inference matching framework providing results that slightly decrease with $\text{MAP}_{log}$ = 0.802 while the results of the edit distance and normalized edit distance techniques respectively drop by 15.9% and 14.9% with $\text{MAP}_{edit\_dist}$ = 0.634 and $\text{MAP}_{norm\_edit\_dist}$ = 0.667. The average precision of our system is approximately 26.5% higher over the average precision of the edit distance based framework and approximately 20.3% higher over the average precision of the system implementing the normalized edit distance.

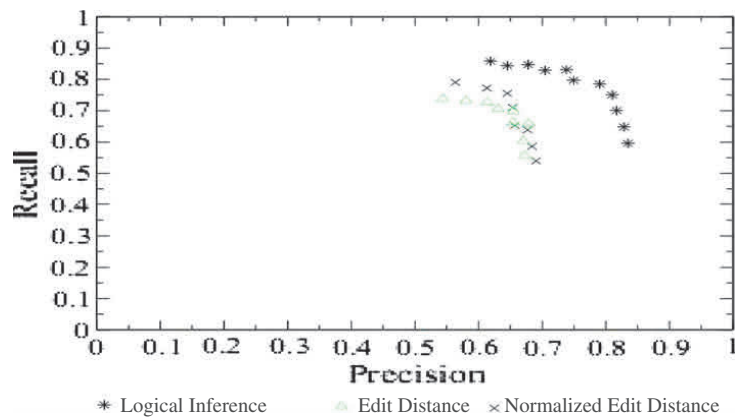

Fig. 9. Precision-recall with short length queries.

## 9. Conclusion and future works

We have presented in this paper a robust architecture integrating signature based indexing and a logical inference based matching and scoring framework to detect near-duplicate videos. In particular, it makes it possible to handle edited and short-length query videos. As far as indexing is concerned, the strength of our proposal relies on the coupling of temporal and perceptual visual (color and texture) features within a unified structure. Regarding the detection process, the matching and scoring framework is based on the computation of the exhaustivity and specificity functions. The latter are respectively based on an *N*-gram sliding window technique for scoring and the computation of path lengths in lattices processing texture and color signatures. Experimental results on two large video corpora show the effectiveness of our approach compared to two dynamic programming-based systems: one relies on edit distance computation and the second implements a normalized edit distance.

Our framework can be potentially used in many commercial applications: commercial and broadcast-stream monitoring, detection of edited contents for regulation authorities. As a research perspective, we will study the integration of audio/speech features within our framework.

## References

Arman, F., Hsu, A., & Chiu, M-Y. (1994). Image processing on encoded video sequences. *Multimedia Systems, 1*(5), 211–219.
Arslan, A., & Egecioglu, O. (1999). An efficient uniform-cost normalized edit distance algorithm. In *Proceedings of SPIRE*.
Berlin, B., & Kay, P. (1991). *Basic color terms: Their universality and evolution*. UC Press.
Bertini, M., Bimbo, D., & Nunziati, W. (2006). Video clip matching using mpeg-7 descriptor and edit distance. In *Proceedings of CIVR* (pp. 133–142).
Bhushan, N., & Al (1997). The texture lexicon: Understanding the categorization of visual texture terms and their relationship to texture images. *Cognitive Science, 21*(2), 219–246.
Cheung, S., & Zakhor, A. (2000a). Efficient video similarity measurement and search. In *Proceedings of ICIP* (pp. 85–88).
Cheung, S., & Zakhor, A. (2000b). Estimation of web video multiplicity. In *Proceedings of SPIE internet imaging* (pp. 34–46).
Eickeler, S., & Hesseler, W. (2006). MPEG-2 compressed domain algorithms for video analysis. *EURASIP Journal on Applied Signal Processing*, 1–11.
Gong, Y., & Al (1996). Image indexing and retrieval based on color histograms. In *Multimedia tools and app. II* (pp. 133–156).
Hampapur, A., & Bolle, R. (2002). VideoGREP: Video copy detection using inverted file indices. IBM Technical Report.
Hampapur, A., Hyun, K. H., & Bolle, R. (2001). Comparison of sequence matching techniques for video copy detection. In *Proceedings of storage and retrieval for media databases* (pp. 194–201).
Hoad, T. C., & Zobel, J. (2003). Fast video matching with signature alignment. In *Proceedings of MIR'03* (pp. 262–269).
Hoad, T. C., & Zobel, J. (2006). Detection of video sequences using compact signatures. In *ACM transactions on information systems* (Vol. 24, No. 1, pp. 1–50).
Joly, A., Frèlicot, C., & Buisson, O. (2003). Robust content-based video copy identification in a large reference database. In *Proceedings of international conference on image and video retrieval*.
Joly, A., Buisson, O., & Frélicot, C. (2007). Content-based copy detection using distortion-based probabilistic similarity search. In *IEEE transactions on multimedia*.
Ke, Y., Sukthankar, R., & Huston, L. (2004). An efficient parts-based near-duplicate and sub-image retrieval system. *In Proceedings of ACM MM* (pp. 869–876).
Lowe, D. (2003). Distinctive image features from scale-invariant keypoints. In *IJCV* (Vol. 20, pp. 91–110).
Mohan, R. (1998). Video sequence matching. In *Proceedings of the international conference on acoustics, speech and signal processing (ICASSP)*.
Vidal, E., Marzal, A., & Aibar, P. (1995). Fast computation of normalized edit distances. In *IEEE transactions on pattern analysis and machine intelligence* (pp. 899–902).
Wu, X., Hauptmann, A. G., & Ngo, C.-W. (2007). Practical elimination of near-duplicates from web video search. In *Proceedings of ACM MM* (pp. 218–227).
Yeh, M.-C., & Cheng, K.-T. (2008). A string matching approach for visual retrieval and classification. In *Proceedings of ACM MIR* (pp. 52–58).
Zabih, R., Miller, J., & Mai, K. (1995). Feature-based algorithms for detecting and classifying scene breaks. In *Proceedings of the 4th ACM international conference on multimedia*.
Zhang, D., & Chang, S.-F. (2004). Detecting image near-duplicate by stochastic attributed relational graph matching with learning. In *Proceedings of ACM MM* (pp. 877–884).
Zhang, J., Kankanhalli, A., & Smoliar, S. W. (1993). Automatic partitioning of full motion video. *Multimedia Systems, 1*(1), 10–28.
Zhao, W.-L., Ngo, C.-W., Tan, H.-K., & Wu, X. (2007). Near-duplicate keyframe identification with interest point matching and pattern learning. In *IEEE trans. on MM* (pp. 1037–1048).
Zhou, J., & Zhang, X.-P. (2005). Automatic identification of digital video based on shot level sequence matching. In *Proceedings of ACM MM* (pp. 515–518).